\documentclass[pre,showpacs,preprintnumbers,amsmath,amssymb]{revtex4}
\usepackage{graphicx}
\usepackage{bm}
\begin{document}

\title{
Spatial persistence and survival probabilities for fluctuating interfaces}

\author{ M. Constantin}
\affiliation{
Condensed Matter Theory Center,
Department of Physics, University of Maryland, College Park, Maryland
20742-4111}
\affiliation{Materials Research Science and Engineering Center,
Department of Physics, University of Maryland, College Park, MD 20742-4111
}
\author{ S. \surname{Das Sarma}}
\affiliation{
Condensed Matter Theory Center,
Department of Physics, University of Maryland, College Park, Maryland
20742-4111}
\author{ C. Dasgupta}
\altaffiliation{Permanent address: Department of Physics, Indian
Institute of Science, Bangalore 560012, India
}
\affiliation{
Condensed Matter Theory Center,
Department of Physics, University of Maryland, College Park, Maryland
20742-4111}

\begin{abstract}
We report the results of numerical investigations of the steady-state
(SS) and finite-initial-conditions (FIC) spatial persistence 
and survival probabilities for (1+1)--dimensional interfaces with
dynamics governed by the nonlinear Kardar--Parisi--Zhang (KPZ)
equation and the linear Edwards--Wilkinson (EW) equation with both
white (uncorrelated) and colored (spatially correlated) noise. 
We study the effects of a finite sampling distance on the measured 
spatial persistence probability and show that both SS and FIC persistence 
probabilities exhibit simple scaling behavior as a function of the 
system size and the sampling distance. Analytical expressions for the 
exponents associated with the power-law decay of SS and FIC 
spatial persistence probabilities of the EW equation with power-law 
correlated noise are established and numerically verified.
\end{abstract}

\pacs{68.35.Ja, 68.37.Ef, 05.20.-y, 0.5.40.-a}

\maketitle
\date{today}

\section{Introduction}
\label{intro}

The concept of temporal persistence \cite{1}, which is closely
related to first-passage statistics, has been used recently
to study various non-Markovian stochastic processes both theoretically
\cite{oldth,1th} and experimentally \cite{1exp,exp_dan,exp_dan2,exp_px}.
Another quantity of interest in the study of the statistics of spatially
extended systems is its natural analog, the {\it spatial persistence
probability}. This idea has been investigated theoretically \cite{maj1}
in the context of $(d+1)$--dimensional Gaussian interfaces with dynamics
described by linear Langevin equations, where the variable undergoing
stochastic evolution is the height $h(x,t)$ of the interfacial sites
($x$ is the lateral position along the interface and $t$ is the time).
The spatial persistence probability of fluctuating interfaces,
denoted by $P(x_{0},x_{0}+x)$, is simply the probability that
the height of a steady-state interface configuration, measured
at a fixed time $t_0$, {\it does not} return to its ``original'' 
value $h(x_0,t_0)$ at the initial point $x_0$ within a distance 
$x$ measured from $x_{0}$ along the interface. In the long-time, 
steady-state limit, the spatial persistence probability 
$P(x_{0},x_{0}+x)$, which depends only on $x$ for a translationally 
invariant interface, has been shown \cite{maj1} to exhibit a power-law 
decay, $P(x_{0},x_{0}+x)\sim x^{-\theta}$. One of the interesting 
results reported in Ref.\cite{maj1} is that the spatial persistence 
exponent $\theta$ can take two values determined by the initial 
conditions or selection rules imposed on the starting point 
$x_0$: ~1) $\theta=\theta_{SS}$, the ``steady state'' (SS) persistence
exponent if $x_{0}$ is sampled uniformly from {\it all} the sites 
of a steady-state configuration; and~~2) $\theta=\theta_{FIC}$, 
the so-called {\it finite-initial-conditions} (FIC) persistence 
exponent if the sampling of $x_0$ is performed from a {\it subset}
of steady-state sites where the height variable and its spatial
derivatives are {\it finite}. The spatial persistence probabilities
obtained for these two different ways of sampling the initial point
are denoted by $P_{SS}(x_{0},x_{0}+x)$ and
$P_{FIC}(x_{0},x_{0}+x)$, respectively.

The values of the exponents $\theta_{SS}$ and $\theta_{FIC}$
for interfaces with dynamics described by a class of linear Langevin 
equations have been determined in Ref.~\cite{maj1} using a
mapping between the spatial statistical properties of the interface 
in the steady state and the temporal properties of
stochastic processes described by a generalized random-walk equation.
It turns out that for these systems, $\theta_{SS}$ is equal to either
$3/2-n$ for $1/2<n<3/2$ or $0$ for
$n>3/2$, where $n=(z-d+1)/2$, $d$ is the spatial dimension, 
and $z$ is the standard dynamical exponent of the underlying Langevin 
equation. The FIC spatial persistence exponent is found to have the
value $\theta_{FIC} = \theta(n)$, where $\theta(n)$ is a temporal 
persistence exponent for the generalized random walk problem to 
which the spatial statistics of the interface is mapped. 
Two exact results for $\theta(n)$ are available in the literature:
$\theta(n=1)=1/2$, corresponding to the classical Brownian motion
\cite{brownian} and $\theta(n=2)=1/4$, corresponding to the
random acceleration problem \cite{random_acc}.

Very recently, experimental measurements of the spatial persistence
probability have been performed \cite{exp_px} for a system 
(combustion fronts in paper) that is believed to
belong to the Kardar--Parisi--Zhang (KPZ)~\cite{kpz} universality class. 
However, the FIC spatial persistence probability is not investigated 
at all in this work. Instead, the authors analyze a ``transient'' 
spatial persistence (i.e., the probability is measured by sampling 
over all the sites of a transient interfacial profile obtained before 
the steady state is reached). This transient spatial persistence is
completely different from the FIC spatial persistence
which is measured in the steady-state regime by sampling a special class
of initial sites. As a consequence, additional study is required 
in order to understand the experimental and numerical possibilities 
for measuring $P_{FIC}$ and its associated nontrivial exponent 
$\theta_{FIC}$. 

In this paper, we present the results of a detailed numerical study of 
spatial persistence in a class of one-dimensional models of fluctuating
interfaces. Our interest in analyzing the spatial persistence of fluctuating
interfaces is motivated to a large extent by their important 
(and far from completely understood) role in the rapidly developing 
field of nanotechnology where the desired stability of nanodevices requires 
understanding and controlling thermal interfacial fluctuations. In
this context, the study of first-passage statistics in general, or of
the persistence probability (both spatial and temporal) \cite{maj1,1th}
in particular, turns out to be a very useful approach. To address
this problem we consider stochastic interfaces with dynamics governed by the
Edwards--Wilkinson (EW)~\cite{ew} and KPZ equations. For the EW equation,
we consider both white noise (uncorrelated in both space and time)
and ``colored''noise that is correlated in space but uncorrelated in time.
The effect of noise in spatially distributed systems is an interesting 
problem by itself and has been widely studied \cite{noise}. In this
paper, we investigate the effects of noise statistics on the spatial 
structure of fluctuating interfaces using the conceptual tool of spatial persistence
probability. Using the isomorphic mapping procedure of Ref.~\cite{maj1}, we
derive exact analytical results for the spatial persistence exponents
of $(d+1)$--dimensional EW interfaces driven by power-law correlated noise.
We then compare our analytical results with those obtained from 
numerical integrations of the corresponding stochastic equations. The use of
power-law correlated noise in the EW equation allows us to explore the
situation where the two spatial persistence exponents $\theta_{SS}$ and
$\theta_{FIC}$ are different.

Our numerical study also provide a characterization of the scaling 
behavior of spatial persistence probabilities as functions of the 
system size. Information about the system-size dependence of persistence 
probabilities is necessary for extracting the
persistence exponents from experimental and numerical data.
In studies of the scaling behavior of spatial persistence probabilities,
one has to consider another important length scale that always
appears in practical measurements: this is the {\it sampling distance} 
$\delta x$ which represents the ``nearest-neighbor''spacing 
of the uniform grid of spatial points where the height variable
$h(x,t_0)$ is measured at a fixed time $t_0$. The sampling distance
$\delta x$ is the spatial analog of the ``sampling time''~\cite{maj2,surv}
that represents the time-interval between two successive measurements 
of the height at a fixed position in experimental and computational 
studies of temporal persistence. Once the effect of a finite
$\delta x$ on the measured spatial persistence is understood, one
can relate correctly the experimental and numerical results to the
theoretical predictions. Our study shows that the spatial persistence
probabilities (both SS and FIC) exhibit simple scaling behavior as 
functions of the system size and the sampling distance.

In addition to the temporal persistence probability, the temporal survival
probability \cite{exp_dan,surv} has been shown recently to represent
an alternative valuable statistical tool for investigations of
first-passage properties of spatially extended systems with stochastic 
evolution. In the context of interface dynamics, the temporal survival 
probability is defined as the probability that the height of the interface
at a fixed position does not cross its {\it time-averaged} value over 
time $t$. In contrast to the power-law behavior of the temporal
persistence probability (which, we recall, measures the probability 
of not returning to the initial position), the temporal
survival probability exhibits an exponential decay at long times,
providing information about the underlying physical mechanisms and their
associated time scales \cite{surv}. In this study, we make the first attempt
to analyze the behavior of the {\it spatial survival probability}, 
$S(x_0,x_0+x)$, defined as the probability of the interface height 
between points $x_0$ (which is an arbitrarily chosen initial position) 
and $x_0+x$ not reaching the average level $\langle h \rangle$ (rather 
than the original value $h(x_0,t_0)$). We present numerical results 
for $S(x_0,x_0+x)$ that show that its spatial behavior in the SS 
regime is neither power-law, nor exponential, while in the FIC 
regime, it becomes very similar to the spatial persistence 
probability, $P_{FIC}(x_0,x_0+x)$.

The paper is organized as follows. In Sec.~\ref{models}, we define
the models studied in this paper, review existing analytical results
about their spatial persistence properties,
and present new analytical expressions for the spatial persistence
exponents for EW interfaces with colored noise in
arbitrary spatial dimension. In Sec.~\ref{methods}, we describe the
numerical methods used in our study and discuss how the
spatial persistence and survival probabilities are measured 
in our numerical simulations. The results obtained in
our (1+1)--dimensional numerical investigations are described 
in detail and discussed in Sec.~\ref{sim}, for both discrete 
stochastic solid-on-solid models (Sec.~\ref{sim}A) and the spatially 
discretized EW equation with colored noise (Sec.~\ref{sim}B).
Sec.~\ref{concl} contains a summary of the main results and a few 
concluding remarks.

\section{Stochastic equations for fluctuating interfaces}
\label{models}

We have performed a detailed numerical study of the spatial
persistence of (1+1)--dimensional fluctuating interfaces where the
dynamics is described by the well known EW equation
\begin{equation}
\frac {\partial h(x, t)} {\partial t} = \nabla^{2} h(x,t) + \eta(x, t),
\label{EW}
\end{equation}
or alternatively by the KPZ equation
\begin{equation}
\frac {\partial h(x, t)} {\partial t} = \nabla^{2} h(x,t) +
(\nabla h(x,t))^{2}+ \eta(x, t),
\label{KPZ}
\end{equation}
\noindent where $\nabla$ and $\nabla^{2}$ refer to spatial derivatives
with respect to $x$, and $\eta(x,t)$ with $\langle \eta(x, t)\eta(x^{'},t^{'})
\rangle \propto \delta(x-x^{'}) \delta(t-t^{'})$ is the usual
uncorrelated random Gaussian noise. The dynamical exponent for 
Eq.(\ref{EW}) is 
$z=2$, and since $d=1$ in our study, the variable $n$ defined in 
Sec.~\ref{intro} is equal to 1. So,
we expect both $\theta_{SS}$ and $\theta_{FIC}$ for this system to be equal to
$1/2$~\cite{maj1}. Although the KPZ equation is nonlinear,
characterized by $z=3/2$, it is well-known that in the long
time limit, the probability distribution of the stochastic height
variable $h(x,t)$ in this equation is the same as that in the EW equation
(i.e. $P(h) \sim \exp [~-\int dx (\nabla h)^{2} ~]$ ) in (1+1) dimensions.
The static roughness exponent, $\alpha$, is the same ($\alpha=1/2$)
for both cases. The 1+1--dimensional KPZ model differs from
the EW model in the {\it transient} scaling regime where the
interfacial roughness grows as a power-law in time, but this temporal
regime is not involved in the calculation of the spatial persistence
probabilities, as explained in Sec.~\ref{intro}. As a consequence,
the steady-state spatial properties of (1+1)--dimensional interfaces 
governed by Eq.~(\ref{KPZ}) can be mapped, as for Eq.~(\ref{EW}), into a
stochastic process with $n=1$. So, the expected values of
$\theta_{SS}$ and $\theta_{FIC}$ for the (1+1)--dimensional KPZ
universality class are also equal to $1/2$. Thus, studies of 
(1+1)--dimensional KPZ and EW interfaces do not bring out the interesting
possibility of different values for the spatial persistence exponents
$\theta_{SS}$ and $\theta_{FIC}$.

To examine the theoretical prediction~\cite{maj1} of a possible difference
between the values of $\theta_{SS}$ and $\theta_{FIC}$, we 
consider the case when the interface dynamics is governed by a EW-type
equation with long-range spatial correlations in the noise. Specifically,
we consider Eq.~(\ref{EW}) with Gaussian colored noise \cite{cn} with
variance given by
\begin{equation}
\langle \eta_c (x, t)~\eta_c (x^{\prime},t^{'}) \rangle =g_{\rho}(x-x^{\prime})
\delta(t-t^{\prime}),
\label{coln}
\end{equation}
\noindent where $0 \le \rho < 1/2$ is a parameter that characterizes the spatial
correlation of the noise, and 
\begin{equation}
g_{\rho}(x-x^{\prime})=\left \{\begin{array}{rcl} &|x-x^{\prime}|^{2 \rho -1}&
\mbox{if}~ ~ ~|x-x^{\prime}| \ne 0\\
&g_{\rho}(0)& \mbox{if} ~ ~ ~x=x^{\prime} \end{array} \right.
\label{g}
\end{equation}
\noindent We have chosen $g_{\rho}(0)$ as in Ref.~\cite{cn}
(i.e. $g_{\rho}(0)=1/\rho (1/2)^{2 \rho}$). As discussed below,
the SS and FIC spatial persistence exponents for (1+1)--dimensional 
interfaces described by the EW equation with this 
kind of colored noise are expected to be different 
from one another. This system, thus, provides an opportunity to examine in
detail the role of the choice of the initial points in determining the
form of the decay of the spatial persistence probability.

By applying the isomorphic mapping recipe of Ref.~\cite{maj1} to the
$(d+1)$--dimensional version of Eq.~(\ref{EW}) with colored noise $\eta_c$
whose statistics is defined by 
Eqs.~(\ref{coln}) and (\ref{g}), we obtain the result 
$n=(z-d+1)/2+\rho$ with $z=2$, implying the following analytical
expressions for the spatial persistence exponents:
\begin{equation}
\theta_{SS}=\frac{d}{2}-\rho
\label{expn1}
\end{equation}
and
\begin{equation}
\theta_{FIC}=\theta \left( \frac{3-d}{2}+\rho \right).
\label{expn2}
\end{equation}
Thus, the value of $\theta_{SS}$ is completely determined by the noise
correlation parameter $\rho$. However, based on the range of
values for $\rho$, we can only infer that $\theta_{FIC}$ varies 
(presumably in a continuous manner) between $\theta(\frac{3-d}{2})$ and
$\theta(\frac{4-d}{2})$ as the parameter $\rho$ is increased from 0 to
1/2. For $d=1$, this implies a change from the value $\theta(1) = 1/2$ to
$\theta(3/2)$, expected to lie between 1/2 and $\theta(2)=1/4$, as $\rho$
changes from 0 to 1/2. Since the value of $\theta_{SS}$ for $d=1$ goes to
0 as $\rho$ approaches the value 1/2, it is clear that the values of the
two spatial persistence exponents must be different for a general value
of $\rho$ in the range [0,1/2). This difference would be small for 
$\rho$ near zero (the two persistence exponents have the same value 
for $\rho=0$), and maximum for $\rho$ near 1/2. Therefore, the model 
with $\rho$ substantially different from zero provides a numerically 
tractable situation where the interesting theoretical prediction of 
the existence of two different nontrivial spatial persistence 
exponents can be tested. We also mention that the usual dynamical 
scaling exponents take the following $\rho$-dependent values in the 
model with colored noise: $\alpha=(2-d+2 \rho)/2$, $\beta=(2-d+2\rho)/4$. 
Thus, the general result~\cite{maj1}, $\theta_{SS} = 1-\alpha$, is 
satisfied for all $d$ and $\rho$.

We have investigated these aspects in a detailed numerical
study of models that belong in the universality classes of
the Langevin  equations of Eqs.~(\ref{EW}) and (\ref{KPZ}).
For Eq.~(\ref{EW}) with uncorrelated white noise, we have used a
discrete stochastic solid-on-solid model (the Family model
\cite{fam}) which is rigorously known to belong to the same
dynamical universality class.
For Eq.~(\ref{KPZ}) with uncorrelated white noise, we have
also used a discrete solid-on-solid model (the Kim--Kosterlitz
model \cite{kk}). Finally, for the EW equation with colored noise, 
the numerical results were obtained from a direct numerical integration
of the spatially discretized stochastic differential equation.

\section{Numerical methods}
\label{methods}
Simulations of the atomistic Family and Kim-Kosterlitz models are 
carried out using the standard Monte Carlo method for implementing 
the stochastic deposition rules of each model. Numerical integration 
of the EW equation with colored noise is performed using the simple 
Euler method~\cite{1th,num_rec}. We solve the (1+1)--dimensional 
Eq.~(\ref{EW}) with spatially long-range correlated noise for the 
real variable $h(x_j,t_n)$, where $t_n=n \Delta t$ ($n=0,1,\ldots$) and
$x_j=j \Delta x$ ($j=0,1,\ldots,L-1$), with periodic boundary
conditions. Here, $\Delta t$ and $\Delta x$ are the spatial and temporal 
grid spacings, respectively.
Using the forward-time centered-space representation \cite{num_rec},
Eq.~(\ref{EW}) becomes:
\begin{equation}
\label{discr_cnEW}
h(x_j,t_{n+1})-h(x_j,t_n)=\Delta t \left[ \frac{h(x_{j+1},t_n)
-2 h(x_j,t_n)+ h(x_{j-1},t_n)}{ (\Delta x)^2}\right] + \sqrt{\Delta t}~
\eta_c(x_j,t_n).
\end{equation}

\noindent We have chosen $\Delta x = 1$ and $\Delta t$ small enough
(i.e. $\Delta t=  0.01$) in order to satisfy the stability criterion
$2 \Delta t/(\Delta x)^2 \le 1$. The spatial correlation of the noise is
given by
\begin{equation}
\label{discr_cn}
\langle \eta_c (x_j, t_n)~\eta_c (x_k, t_m) \rangle =g_{\rho}(x_j-x_k)
\delta_{n,m}
\end{equation}
\noindent with
\begin{equation}
g_{\rho}(x_j-x_k)=\left \{\begin{array}{rcl} &|x_j-x_k|^{2 \rho -1}&
\mbox{if}~ ~ ~1 \le |x_j-x_k| \le \frac{L}{2} \\
&(L-|x_j-x_k|)^{2 \rho -1}&
\mbox{if}~ ~ ~|x_j-x_k| > \frac{L}{2} \\
&g_{\rho}(0)& \mbox{if} ~ ~ ~x_j-x_k=0 \end{array} \right.
\label{discr_g}
\end{equation}
\noindent where $g_{\rho}(0)=1/\rho (1/2)^{2 \rho}$. The
colored noise is generated using the recipe from Ref. \cite{cn}. 
The Fast Fourier
Transform operation that is used in the noise-generation procedure
constrains the system size to be an integral power of 2. Due to the
use of periodic boundary conditions (which are also imposed on the noise 
correlation function, see Eq.~(\ref{discr_g})), the range of $x$ over which 
spatial correlations and persistence properties 
are meaningfully measured is of the order of $L/2$.

The SS spatial persistence probability $P_{SS}(x_{0},x_{0}+x)$
is measured at a fixed time $t_0$ (which is much larger than the 
time $t_{sat} \sim L^z$
required for the interface roughness to saturate) as the probability
that the interface height variable does not cross its value, $h(x_0,t_0)$,
at the initial point $x_0$ as one moves along the interface from the
point $x_0$ to the point $x_0+x$.  
This probability is averaged over all the sites in a
steady-state configuration and also over many independent realizations
of the stochastic evolution. Thus,
\begin{equation}
P_{SS}(x_{0},x_{0}+x) \equiv \hbox{Prob}~\lbrace ~
\hbox{sign}\,[h(x_{0}+x^{\prime}) - h(x_{0})] = \,\,\hbox{constant},\,\,
~\forall~ 0< x^{\prime} \leq x ~, ~\forall ~ x_{0} \in {\cal S}_{SS}~
\rbrace,
\label{probSS}
\end{equation}
\noindent where $\hbox{sign}\,[y]$ represents the sign 
of the fluctuating quantity $y$, and ${\cal S}_{SS}$ is the ensemble 
containing all the lattice sites in a steady-state configuration. 
The FIC spatial persistence probability $P_{FIC}(x_{0},x_{0}+x)$ is
obtained in a similar manner, except that the
average is performed over a particular subensemble of the steady-state
configuration sites, ${\cal S}_{FIC} \subset {\cal S}_{SS}$,
characterized by {\it finite} values of the height variable and its
spatial derivatives:
\begin{equation}
P_{FIC}(x_{0},x_{0}+x) \equiv \hbox{Prob}~\lbrace
~\hbox{sign}\,[h(x_{0}+x^{\prime}) - h(x_{0})] = \,\,\hbox{constant},\,\,
~\forall~ 0< x^{\prime} \leq x ~, ~\forall ~ x_{0} \in {\cal S}_{FIC}~
\rbrace. 
\label{probFIC}
\end{equation}
Since the persistence probabilities are averaged over the choice of
the initial point $x_0$, we omit writing $x_{0}$ explicitly in
the arguments of $P_{SS}$ and $P_{FIC}$ from now on, while stressing the 
important fact that the ensemble of initial sites used in the
averaging process determines which one of the two persistence 
probabilities is obtained. We consider two different methods for 
measuring $P_{FIC}(x)$, depending on the type of the model (atomistic
solid-on-solid model or spatially discretized Langevin equation) being 
studied. In the former case where the height variables are integers, 
the FIC spatial persistence probability measurement involves
a sampling procedure from the subset of sites characterized by a fixed
integer value of the height (measured from the average, $\langle h \rangle$, of the 
heights of all the sites at time $t_0$) which is substantially smaller than the
typical value of the height fluctuations measured by the saturation width
of the interface profile. In calculations using the direct numerical 
integration technique, the height variable can take any real value. 
So, the probability of finding a fixed value of the stochastic height 
variable is infinitesimally small. For this reason, fixing a reference 
level $H$ and sampling over the sites with $h(x_{0},t_0)=\langle h
\rangle+ H$ is useless. We, therefore, consider in this case 
a continuous interval of height values (symmetric with
respect to the average height $\langle h \rangle$) with width $w$ which is
considerably smaller than the amplitude of the height fluctuations.
The positions characterized by a height variable within this interval
represent the subensemble of lattice positions involved in the sampling
procedure necessary for measuring $P_{FIC}(x)$.

The spatial survival probabilities corresponding to the SS and FIC conditions
are calculated similarly to the corresponding persistence probabilities,
except that the stochastic variable under consideration becomes
$h(x_0+x^{\prime})- \langle h \rangle$. Thus,
\begin{equation}
S_{SS}(x_{0},x_{0}+x) \equiv \hbox{Prob}~\lbrace ~
\hbox{sign}\,[h(x_{0}+x^{\prime}) - \langle h \rangle] =\,\,\hbox{constant}\,\,,
~\forall~ 0\le x^{\prime} \leq x ~, ~\forall ~ x_{0} \in {\cal S}_{SS}~\rbrace,
\label{surv_SS}
\end{equation}
\noindent and
\begin{equation}
S_{FIC}(x_{0},x_{0}+x) \equiv \hbox{Prob}~\lbrace
~\hbox{sign}\,[h(x_{0}+x^{\prime}) - \langle h \rangle] =\,\,\hbox{constant}\,\,
~\forall~ 0 \le x^{\prime} \leq x ~, ~\forall ~ x_{0} \in {\cal S}_{FIC}~\rbrace.
\label{surv_FIC}
\end{equation}

\section{Results and discussions}
\label{sim}
\subsection{Solid-on-solid models}
\label{A}

In the solid-on-solid Family and Kim--Kosterlitz models,
the interface configuration is characterized by a set of integer
height variables $\lbrace h_i\rbrace_{i=1,L}$ corresponding to the
lattice sites $i=1,\ldots ,L$, with periodic boundary conditions.
Since all the measurements of the spatial persistence and survival
probabilities are done in the steady-state regime (i.e. in the regime 
where the interfacial
roughness has reached a time-independent saturation value), we used
relatively small systems with $L \sim 200-3000$ in order to be able to
achieve the the steady state within reasonable simulation times. The
resulting steady-state interfacial profile, corresponding to a final
time $t_0 >> L^z$, is used to compute the spatial persistence and
survival probabilities. The calculation of $P_{SS}(x)$ is relatively
simple: it involves measuring the fraction of initial 
lattice positions (all possible choices of the initial point are allowed)
for which the interface height has
not returned to the height of the initial point (for persistence probability) or
to the average height level $\langle h \rangle$ (for survival probability)
over a distance $x$, averaged over many independent realizations
($\sim 10^3-10^4$) of the steady state configuration. 
Measurements of $P_{FIC}(x)$ or $S_{FIC}(x)$ involve,
in addition to these steps, a preliminary selection of a 
subensemble of lattice sites which are characterized by a 
fixed and small value $H$ of the height measured relative 
to the spatial average. Only the sites that belong to this 
subensemble (i.e. only the sites with $h_i=H+\langle h \rangle$) 
are used as initial points in the FIC measurements.

\begin{figure}
\includegraphics[height=6cm,width=16cm]{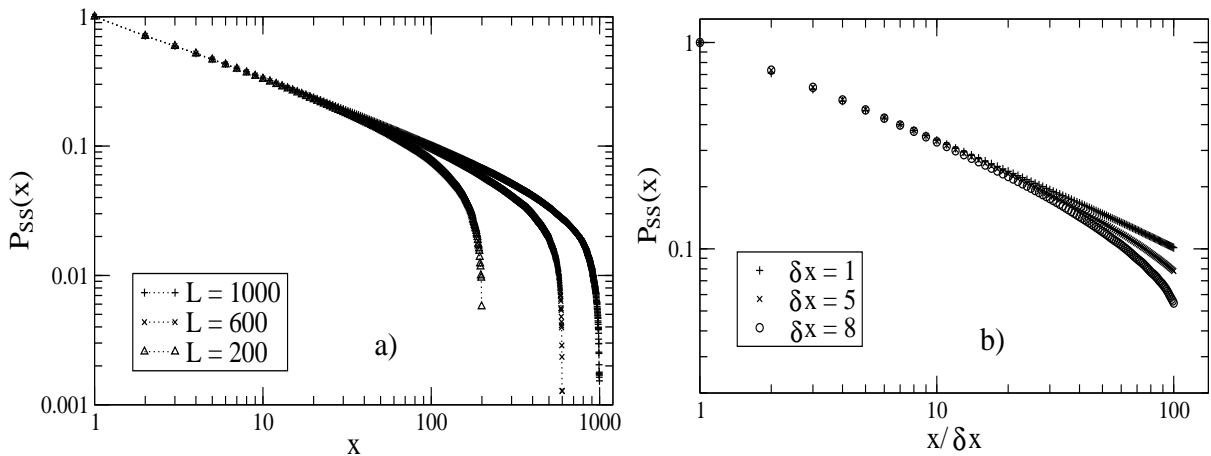} 
\caption{\label{fig1} The steady state spatial persistence
probability, $P_{SS}(x)$, for (1+1)--dimensional EW interfaces with 
white noise, obtained
using the discrete Family model.
Panel (a): Double-log plots of $P_{SS}(x)$ vs $x$ for a fixed sampling distance
$\delta x=1$, using three different values of $L$,
as indicated in the legend.
Panel (b): Double-log plots of $P_{SS}(x)$ vs $x/\delta x$ for a 
fixed system size, 
$L=1000$, and 
three different values of $\delta x$, as indicated in the legend.}
\end{figure}

Two distinct length scales have to be taken into consideration in the
interpretation of the numerical results for the
spatial persistence probability: the size $L$ of the sample used in the 
simulation, and the sampling distance $\delta x$ which denotes the spacing
between two successive points where the height variables are measured in the
calculation of the persistence probability. The minimum value of $\delta x$
is obviously one lattice spacing, but one can use a larger integral value
of $\delta x$ in the calculation of persistence and survival probabilities.
For example, a calculation of the persistence probability with $\delta x = m$
would correspond to checking the heights of only the sites with index 
$i_0+jm$, where $i_0$ is the index of the initial site and $j=1,2,\ldots$.
While the importance of $L$ in the measurement of $P(x)$ is obvious 
(it sets the maximum distance for which $P(x)$ can be meaningfully measured), 
the effect of $\delta x$ is rather intricate and has to be carefully investigated.
In Fig.~\ref{fig1}a we start to analyze these effects by looking
at $P_{SS}(x)$ for EW-type interfaces. We note that when $P_{SS}(x)$ is
measured in systems with different sizes, using the smallest
possible value for $\delta x$ (i.e. $\delta x=1$), the exponent
associated with the power-law decay of the persistence probability does
not change, but there is an abrupt downward departure from a power-law behavior
near $x=L/2$. It is not difficult to understand this behavior qualitatively:
as discussed earlier, measurements of spatial correlations and persistence
probabilities in a finite system of size $L$ with periodic boundary
conditions are meaningful only for distances smaller than $L/2$. 
In Fig.~\ref{fig1}b, we have shown the results for $P_{SS}(x)$
when $L$ remains fixed and $\delta x$ is varied. Since the the persistence 
probability is, by definition, equal to unity for $x=\delta x$ (see 
Eq.~(\ref{probSS})), we have plotted $P_{SS}$ 
as a function of $x/\delta x$ in this figure to ensure that the plots for
different values of $\delta x$ coincide for small values of the $x$-coordinate.
The plots for different $\delta x$ are found to splay away from each other
at large values of $x/\delta x$, with the plots for larger $\delta x$ 
exhibiting more pronounced downward bending. Again, the reason for this
behavior is 
qualitatively clear: since a double-log plot of $P_{SS}(x)$ vs $x$ begins to
deviate substantially from linearity as $x$ approaches $L/2$ 
(see Fig.~\ref{fig1}a), the downward bending of the plots in Fig.~\ref{fig1}b
(which are all for a fixed value of $L$) occurs at a smaller value of
$x/\delta x$ for larger $\delta x$.  
A more detailed scaling analysis of the dependence of the persistence 
probabilities on $x$ and $\delta x$ is described below.

\begin{figure}
\includegraphics[height=11cm,width=16cm]{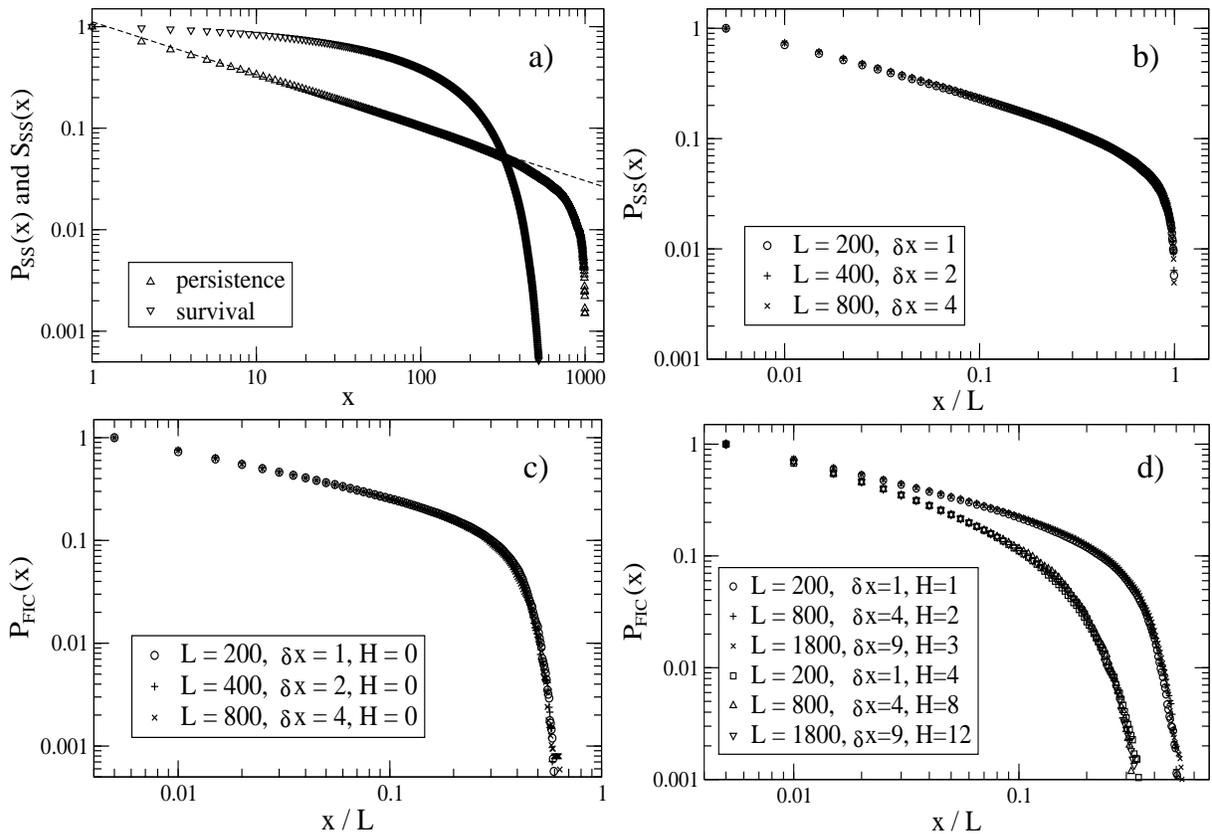} 
\caption{\label{fig2} The spatial persistence
probabilities, $P_{SS}(x)$ and $P_{FIC}(x)$, and the spatial survival
probability, $S_{SS}(x)$, obtained from simulations of the
Family model in (1+1) dimensions. In panels (a) and (b) we show the 
data for $P_{SS}(x)$ and $S_{SS}(x)$, while in panels (c) and (d) 
we display the data for $P_{FIC}(x)$. 
Panel (a): $P_{SS}(x)$ and $S_{SS}(x)$ for $L=1000$, $\delta x=1$.
The dashed line represents the best fit of the $P_{SS}(x)$
data to a power-law form.
Panel (b): Finite-size scaling of $P_{SS}(x,L,\delta x)$. Three
probability curves are obtained for three different sample sizes with the same
value for the ratio $\delta x/L = 1/200$.
Panel (c): Scaling of $P_{FIC}(x,L,\delta x,H)$ for the same values of $L$ and
$\delta x$ as in panel (b). $P_{FIC}$ is calculated by 
sampling over lattice sites with $H=0$.
Panel (d): Scaling of $P_{FIC}(x,L,\delta x,H)$ for three different sample
sizes with the same value for the ratio $\delta x/L$, sampling over
two subsets of lattice sites with the same value of $H/L^{\alpha}$
($\alpha=0.5$): $1/\sqrt{200}$ (upper plot) and $4/\sqrt{200}$ (lower plot).
}
\end{figure}

\begin{figure}
\includegraphics[height=11cm,width=16cm]{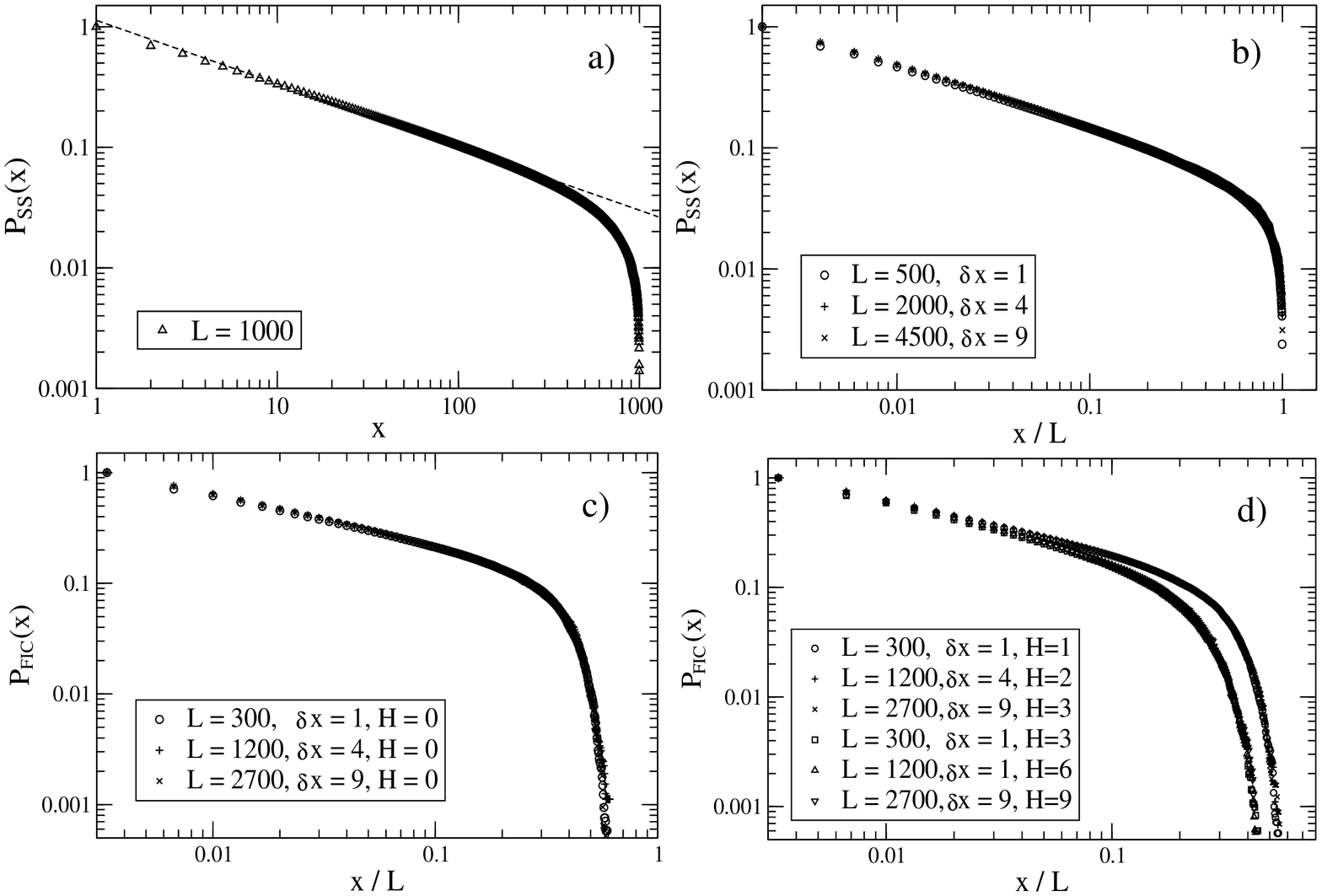} 
\caption{\label{fig3} The spatial persistence
probabilities, $P_{SS}(x)$ and $P_{FIC}(x)$, for the (1+1)-dimensional
Kim-Kosterlitz model which is in the KPZ universality class.
As in Fig.~\ref{fig2}, in panels (a) and (b) we show the data for
$P_{SS}(x)$. Panels (c) and (d) display the data for $P_{FIC}(x)$.
Panel (a): $P_{SS}(x)$ for $L=1000$, $\delta x=1$.
Panel (b): Finite-size scaling of $P_{SS}(x,L,\delta x)$. Three
probability curves are obtained for three different sample sizes with the same
value for the ratio $\delta x/L=1/500$.
Panel (c): Scaling of $P_{FIC}(x,L,\delta x,H)$, obtained by sampling 
over the lattice sites with $H=0$, for three different values 
(same as those in panel (b)) of $L$ and $\delta x$.
Panel (d): Scaling of $P_{FIC}(x,L,\delta x,H)$ for three different sample
sizes with the same value for the ratio $\delta x/L$, sampling over
two subsets of lattice sites with the same value of $H/L^{\alpha}$
($\alpha=0.5$): $1/\sqrt{300}$ (upper plot) and $3/\sqrt{300}$ (lower plot).
}
\end{figure}

In Fig.~\ref{fig2} we show the results for spatial persistence and survival
probabilities for the discrete Family model. It is obvious
from the plots that the spatial persistence probabilities $P_{SS}(x)$ (panel (a))
and $P_{FIC}(x)$ (panel (c)) 
exhibit power-law decays over an extended range of $x$ values.
The abrupt decay to zero near $x=L/2$ is due, as discussed above,
to finite size effects. The spatial persistence exponents are extracted
from the power-law fits shown in the log-log plots as dashed straight lines.
We find that $\theta_{SS}\simeq 0.51$, in good agreement with the expected
value $1/2$. However, it is clear that the steady state survival
probability $S_{SS}(x)$, shown in Fig.~\ref{fig2}a, does not exhibit a
power-law behavior. This is similar to the qualitative behavior of the 
{\it temporal} survival probability in the steady state of the 
Family model~\cite{surv}. 

We now return to the dependence of the persistence probabilities on the
sample size $L$ and the sampling distance $\delta x$. Since $L$ and $\delta x$
are the only two length scales in the problem (the lattice parameter serves
as the unit of length), it is reasonable to expect~\cite{surv} that the
persistence probabilities would be functions of the (dimensionless) scaling
variables $x/L$ and $\delta x/L$. If this is true, then plots of $P$ vs.
$x/L$ for different sample sizes should show a scaling collapse if the
ratio $\delta x/L$ is kept constant. A similar scaling behavior of the
temporal survival probability as functions of $L$ and the sampling time
$\delta t$ (in that case, the scaling variables are $t/L^z$ and $\delta t/L^z$)
was found in Ref.~\cite{surv}.
As indicated in panels(b-d) of Fig.~\ref{fig2}, we
have used various values for the sampling distance $\delta x$ in the measurement
of $P_{SS}(x)$ and $P_{FIC}(x)$. We observe that when the sampling distance
is increased in proportion to the system size (so that $\delta x/L$ is held
fixed), all the $P_{SS}(x)$ curves collapse when
plotted vs. $x/L$ (see panel (b)). This confirms that the scaling
form of the steady state persistence probability is:
\begin{equation}
P_{SS}(x,L,\delta x) = f_{1}(x/L,~ \delta x/L),
\label{f1}
\end{equation}
\noindent where the function $f_{1}(x_{1},x_{2})$ shows a power-law decay 
with exponent $\theta_{SS}$ as a function of $x_1$ for small values of 
$x_1$ and $x_2 \ll 1$. 

Let us turn our attention to $P_{FIC}(x)$. In the data shown in panel (c) of
Fig.~\ref{fig2}, we have chosen the subensemble ${\cal S}_{FIC}$ of sampling
positions to contain only the lattice sites whose height $h_i$ is equal to the
average value $\langle h \rangle$ (i.e. $H=0$). Obviously, in this
case the definitions for persistence and survival probabilities 
become identical, since the probability that the height variable
does not return to the original value (i.e. $h_i=\langle h \rangle$) is precisely
the probability that the height variable does not reach the average level
$\langle h \rangle$. We find that $\theta_{FIC} \simeq 0.48$ using a system with
$L=1000$ and $\delta x=1$ and considering the subensemble of sites with $H=0$.
We note that a remarkable collapse of $P_{FIC}(x)$ vs. $x/L$ curves for 
different values of $L$ is again obtained when $\delta x$ 
is adjusted to be proportional to the system size $L$, as 
shown in panel (c). More interestingly, we observe
that fixing the level $H$ to a nonzero value introduces a ``height''
scale in the problem that is related to the steady-state
value of the interface width.  Since this width is proportional to 
$L^{\alpha}$, where $\alpha$ is the roughness exponent, we expect 
the dependence of $P_{FIC}$ on $H$ for nonzero values of $H$ to 
be described by the scaling variable $H/L^\alpha$. We observe 
that if the level $H$ is chosen to be proportional to 
$L^{\alpha}$, then the calculated values of $P_{FIC}$ for
different sample sizes, obtained using values of $\delta x$ such that
the ratio $\delta x/L$ is also held constant, exhibit a perfect 
scaling collapse, as shown in panel (d) of Fig.~\ref{fig2}.
This observation leads us to the conclusion that the
scaling form of the FIC persistence probability with nonzero values
of the level $H$ is:
\begin{equation}
P_{FIC}(x,L,\delta x,H) = f_{2}(x/L,~ \delta x/L,~H/L^{\alpha}),
\label{f2}
\end{equation}
\noindent where $f_{2}(x_{1},x_{2},x_{3})$ exhibits a power-law behavior
with exponent $\theta_{FIC}$ as a function of $x_1$ for small $x_1$ 
if $x_2 \ll 1$ and  $x_3 \to 0$.
As the value of $x_{3}$ is increased, the range of $x_1$ values
over which the power-law behavior is obtained decreases and a more
rapid decay of the probability is noticed.

The predictions concerning the scaling behavior of the spatial persistence
probabilities are confirmed by the results for 
the atomistic Kim--Kosterlitz
model. The same discussion for Fig.~\ref{fig2} applies to Fig.~\ref{fig3} where
we have shown the results for the Kim-Kosterlitz model.
We find that $\theta_{SS}\simeq 0.52$ (see Fig.~\ref{fig3}a), 
in good agreement with the
expected value of 1/2, and also that $\theta_{FIC} \simeq 0.47$, using
a rather small simulation with $L=300$ and $\delta x=1$ and
sampling over the subensemble of sites with height at the average
level (see Fig.~\ref{fig3}c). As shown in Fig.~\ref{fig3}b, the SS 
persistence probability obeys the scaling form of Eq.~(\ref{f1}).
In Fig.~\ref{fig3}d, we display the results
for the measured $P_{FIC}$ for systems with different
sizes and sampling distances such that $\delta x/L$ remains constant and
considering two different subsets of sampling sites, each subset being
characterized by a fixed value of $H/L^{\alpha}$. These results are
in perfect agreement with the scaling form of Eq.~(\ref{f2}).

Equations (\ref{f1}) and (\ref{f2}) provide a complete scaling 
description of the SS and FIC persistence probabilities 
for (1+1)--dimensional fluctuating interfaces belonging to two 
different universality classes (i.e. EW and KPZ), modeled
using discrete solid-on-solid models. The associated spatial
persistence exponents $\theta_{SS}$ and $\theta_{FIC}$ are in 
good agreement with the theoretical values \cite{maj1}. However, 
these studies do not illustrate the interesting possibility of a 
dependence of the persistence exponent on the sampling procedure 
used in the selection of the initial sites used in the calculation 
of the persistence probability: the two persistence exponents
$\theta_{SS}$ and $\theta_{FIC}$ have the same value for (1+1)--dimensional
EW and KPZ interfaces. We present and discuss below the results for a model
where these two exponents have different values.

\subsection{EW equation with colored noise}
\label{B}

\begin{figure}
\includegraphics[height=6cm,width=16.5cm]{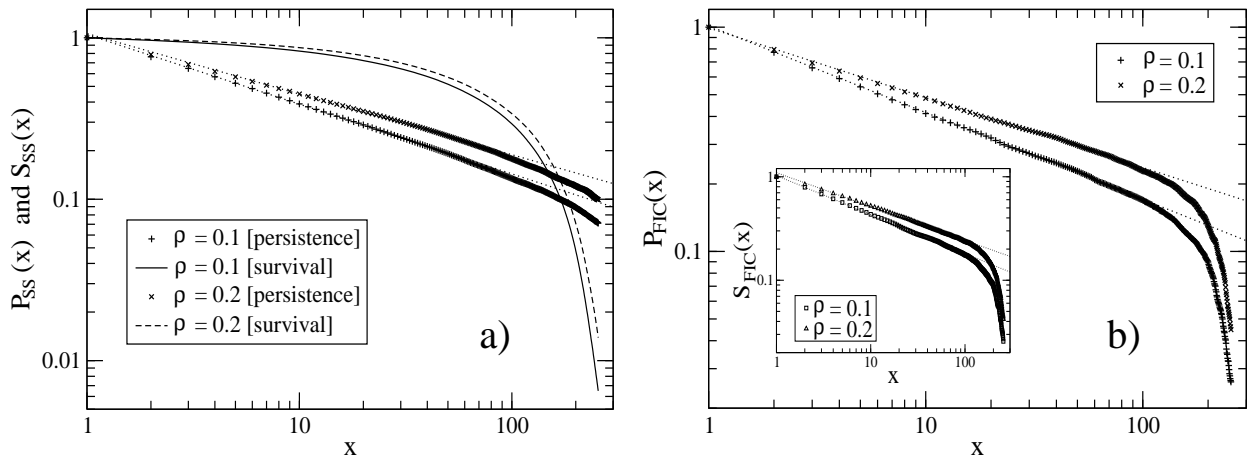} 
\caption{\label{fig4} Spatial persistence and survival probabilities
for the EW equation with spatially correlated noise. 
Panel a): $P_{SS}(x)$ and $S_{SS}(x)$
using a fixed system size $L=2^9$, two values of the noise correlation parameter
($\rho=0.1$ and $0.2$) and sampling distance $\delta x = 1$.
Panel b): $P_{FIC}(x)$ and $S_{FIC}(x)$ (inset), using the same parameters
as in panel a), and sampling initial sites from a band of width $w=0.10$
centered at the average height. The straight lines drawn through the data
points in these double-log plots represent 
power-law fits. }
\end{figure}

\begin{figure}
\includegraphics[height=6cm,width=16.5cm]{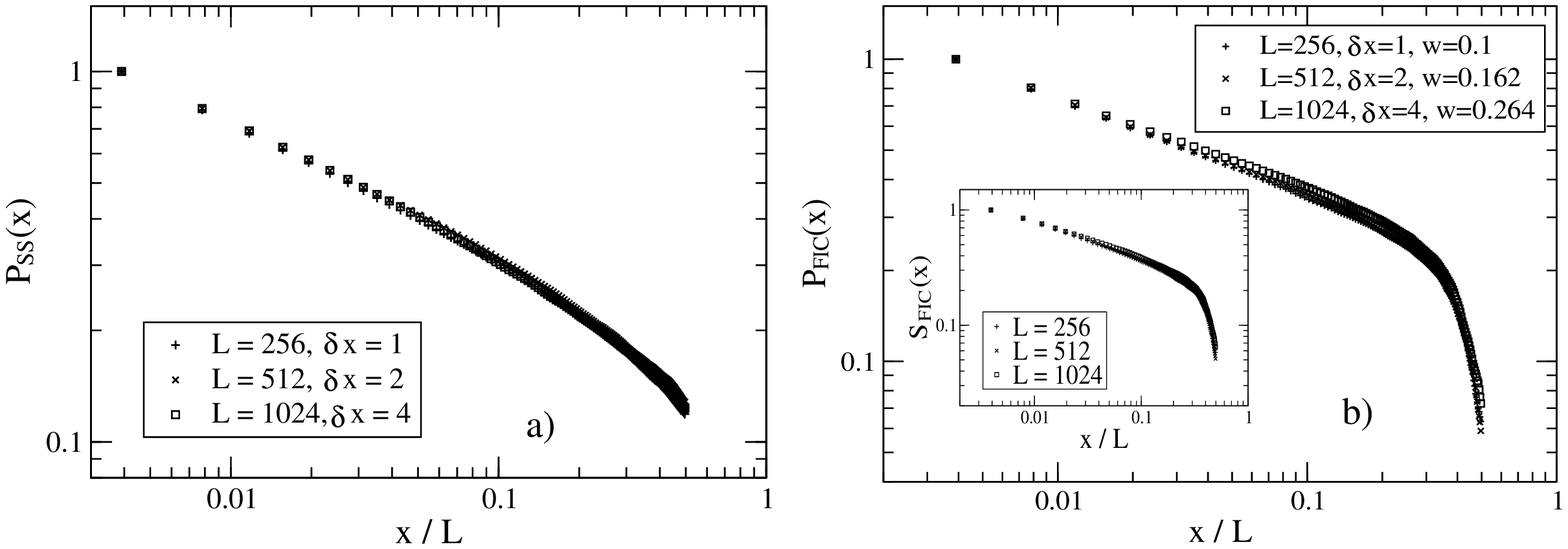} 
\caption{\label{fig5} Finite-size scaling of the persistence
probabilities, $P_{SS}(x)$ and $P_{FIC}(x)$, and the FIC survival probability
$S_{FIC}(x)$ for the EW equation with spatially correlated noise.
The noise correlation parameter is $\rho=0.2$ and the sampling
interval $\delta x$ takes three different values. Panel a): The SS persistence
probability $P_{SS}(x,L,\delta x)$ for three different sample sizes with a
constant ratio $\delta x/L = 1/2^{8}$. Panel b): The FIC persistence probability
$P_{FIC}(x,L,\delta x,w)$ with fixed values of the quantities
$\delta x/L$ ($= 1/2^8$) and $w/L^{\alpha}$ ($=0.1/2^{5.6}$), where $\alpha=0.7$
Inset: Same as in the main figure, but for the FIC survival probability
$S_{FIC}(x,L,\delta x,w)$.}
\end{figure}

In order to measure the spatial persistence and survival 
probabilities in this system, we have applied the steps 
described above on systems of sizes $\sim 2^8-2^{10}$, using 
100--400 independent realizations for averages. While the calculation of
$P_{SS}(x)$ and $S_{SS}(x)$ involves the same method as the one used
in the case of the solid-on-solid models, for measuring $P_{FIC}(x)$ and
$S_{FIC}(x)$ we have selected the subensemble of lattice sites whose
heights $h(x_j,t_0)$ at time $t_0 \gg L^z$ satisfy the condition 
$\langle h \rangle-w/2 \le h(x_j,t_0) \le \langle h \rangle + w/2$, 
where $\langle h \rangle$ is the spatial average 
of the height at time $t_0$. The width $w$ of the sampling 
window has to be chosen to be much smaller than the amplitude 
of the interface fluctuations, but large enough to include a 
relatively large fraction of the total number of sites in order 
to ensure adequate statistics. Under these circumstances we have 
computed the fraction of these selected sites which do not reach
the ``original'' height $h(x_j,t_0)$ (in the case of persistence 
probability) or the average height level $\langle h \rangle$ 
(in the case of survival probability) up to a distance $x$ from 
the point $x_j$. The numerical results for these probabilities,
along with a finite-size scaling analysis of their behavior, are shown
in Figs.~\ref{fig4} and \ref{fig5}.

We find that both SS and FIC spatial persistence probabilities 
for (1+1)--dimensional interfaces described by the EW equation 
with colored noise exhibit the expected power-law behavior as a 
function of $x$, as shown in Fig.~\ref{fig4}, while the SS 
survival probability shows a more complex $x$-dependence 
(see Fig.~\ref{fig4}a). Further work is needed in order
to understand the behavior of $S_{SS}(x)$. When a relatively 
small system with size $L=2^9$ is used, the numerical results 
for the spatial persistence exponents extracted from the power-law
fits shown in Fig.~\ref{fig4} (for $\rho=0.1$, we obtain 
$\theta_{SS}\simeq 0.43$ and $\theta_{FIC}\simeq 0.38$, 
while for $\rho=0.2$, the exponent values are found to be
$\theta_{SS}\simeq 0.37$ and $\theta_{FIC}\simeq 0.31$) appear to be
affected by finite-size effects. Specifically, the values of 
$\theta_{SS}$ extracted from fits to the numerical data are 
systematically larger than the theoretically expected values, 
$\theta_{SS} = 0.4$ for $\rho=0.1$ and 0.3 for $\rho=0.2$ 
(see Eq.~(\ref{expn1})). Similar deviations from the analytical results are
also found for the usual dynamical scaling exponents $\alpha$ and 
$\beta$. We have checked that simulations of larger samples 
bring the measured values of the exponents closer to the expected 
values, but the convergence is rather slow. These finite-size 
effects become more pronounced as the noise correlation parameter 
$\rho$ is increased. In Fig.~\ref{fig4} we show the results
for $\rho=0.1$ and $\rho=0.2$, but we have verified from simulations
with larger values of $\rho$ that the difference between the expected 
and measured values of $\theta_{SS}$ increases as $\rho$ is
increased. This is expected because the spatial correlation of 
the noise falls off more slowly with distance as $\rho$ is increased, 
thereby making finite-size effects more pronounced. 
Another possible source of the discrepancy between the numerical and exact
results for the exponent $\theta_{SS}$ is the spatial discretization used
in the numerical work. The effects of using a finite discretization scale
$\Delta x$ on the observed scaling behavior of continuum growth equations
in the steady state have been studied in Ref.~\cite{buceta} where it was
found that the effective value of the roughness exponent $\alpha$ obtained
from calculations of the local width using a finite $\Delta x$ is smaller
than its actual value. Since $\theta_{SS} = 1-\alpha$, the values of
$\theta_{SS}$ obtained from our calculations with $\Delta x = 1$ are
expected to be larger than their exact values. Our results are consistent
with this expectation.
As shown in the inset of Fig.~\ref{fig4}b, the FIC survival 
probability $S_{FIC}(x)$ behaves similarly to $P_{FIC}(x)$ for 
both $\rho =0.1$ and 0.2, exhibiting a power-law decay with 
an exponent (of $0.38$ and $0.33$ for $\rho=0.1$ and 0.2,
respectively) that is very close to $\theta_{FIC}$.
This is consistent with the expectation that the FIC persistence 
and survival probabilities should become identical as the width 
parameter $w$ used in the selection of initial sites approaches
zero (in this limit, both persistence and survival probabilities 
measure the probability of not crossing the average height).
Finally, we point out that both SS and FIC exponents obtained from
the numerical study exhibit the correct trend, increasing in 
magnitude as $\rho$ decreases. Also, the measured FIC spatial 
persistence exponents satisfy the constraint 
$1/4 < \theta_{FIC} \le 1/2$. Our numerical
results also confirm the interesting theoretical prediction that 
the SS and FIC spatial persistence exponents are different for the 
EW equation with spatially correlated noise.

We have found that the scaling forms of Eqs.~(\ref{f1}) and (\ref{f2})
also provide a correct description of the numerically obtained
persistence and survival probabilities for the EW
equation with spatially correlated noise. This is illustrated 
in Fig.~\ref{fig5}. In Fig.~\ref{fig5}a, we show that the results for
$P_{SS}(x,L,\delta x)$ obtained for different values of $L$ and $\delta x$
fall on the same scaling curve when plotted against $x/L$ if the ratio
$\delta x/L$ is held fixed. This is precisely the behavior predicted by 
Eq.~(\ref{f1}). As shown in Fig.~\ref{fig5}b, the data for 
$P_{FIC}(x,L, \delta x,w)$ also exhibit good finite-size 
scaling collapse if $\delta x$ is varied in proportion to 
$L$ and the width $w$ of the sampling band is increased in 
proportion to $L^{\alpha}$. This is in perfect analogy
with the scaling behavior of the FIC persistence probability for 
the discrete stochastic models discussed in Sec.~\ref{A}, with 
the variable $w$ playing the role of $H$ in Eq.~(\ref{f2}). This 
suggests that the scaling behavior of the FIC persistence probability 
in the continuum EW equation is of the form
\begin{equation}
P_{FIC}(x,L,\delta x,w) = f_{3}(x/L,~ \delta x/L,~w/L^{\alpha}),
\label{f3}
\end{equation}
\noindent where the function $f_3$ has the same characteristics
as $f_2$ in Eq.~(\ref{f2}). A similar scaling description also 
applies to $S_{FIC}(x)$, as shown in the inset of Fig.~\ref{fig5}b. 
This scaling description should be useful in the analysis of 
experimental data on equilibrium step fluctuations 
\cite{exp_dan,exp_dan2} because the images obtained in 
experiments provide the values of a real ``height'' variable 
(position of a step-edge) at discrete intervals of a finite 
sampling distance $\delta x$.

\section{Summary and concluding remarks}
\label{concl}
In this study, we have analyzed the spatial first-passage 
statistics of fluctuating interfaces using the concepts of 
spatial persistence and survival probabilities. Specifically, 
we have presented the results of detailed numerical 
measurements of the SS and FIC spatial persistence probabilities 
for several models of interface fluctuations. Results for
the spatial survival probabilities are also reported. These 
results confirm that the concepts of persistence and survival 
are useful in analyzing the spatial structure of fluctuating 
interfaces. The exponents associated with the power-law decay of the
spatial persistence probabilities as a function of distance $x$ 
are valuable indicators of the universality class of the stochastic 
processes that describe the dynamics of surface fluctuations. 
Our results for these exponents for (1+1)-dimensional interfaces 
in the EW and KPZ universality classes are in good agreement with
the corresponding analytic predictions. We have also obtained analytic
results for the spatial persistence exponents in the (1+1)-dimensional EW
equation with spatially correlated noise, and reported the results of a 
numerical calculation of the persistence and survival probabilities in 
this system. While the numerical results show strong finite-size
effects, the qualitative trends predicted by the analytic treatment are
confirmed in the numerical work. In particular, the numerical results show
evidence for an interesting theoretically predicted difference between the
persistence exponents obtained for two different ways of sampling the 
initial points used in the measurement of the spatial persistence
probability. We also find that the steady-state survival
probability has a complex spatial behavior that requires further 
investigations. In the past, there has been some confusion in the 
literature about the distinction between the persistence and 
survival probabilities~\cite{surv}. Our study shows that these 
two quantities are very different in the SS situation, whereas 
the distinction between them essentially disappears in the FIC
situation.

The numerical results reported here are for models that exhibit ``normal''
scaling behavior with the same local and global scaling properties of
interface fluctuations. There are other models of interface growth and
fluctuations that exhibit ``anomalous'' scaling~\cite{anomsc}, for which
the global and local scaling properties are different. In such models, the
``global'' roughness exponent $\alpha_g$ that describes the dependence of
the interface width in the steady state on the sample size $L$ ($W(t_0,L)
\propto L^{\alpha_g}$ for $t_0 \gg L^z$) is different from the ``local''
exponent $\alpha_l$ that describes the $x$-dependence of the
height-difference correlation function $g(x) \equiv \langle [h(x+x_0,t_0)
- h(x_0,t_0)]^2 \rangle ^{1/2}$ in the steady state ($t_0 \gg L^z$) for
small $x$ ($g(x) \propto x^{\alpha_l}$ for $x \ll L$). The exponent
$\alpha_g$ is greater than unity (the steady-state interface is
``super-rough'') in such cases, whereas the local exponent $\alpha_l$ is
always less than or equal to unity. It is interesting to enquire about the
behavior of the spatial persistence probabilities in such models. The
numerical results reported in the preceding sections show that the
steady-state persistence probability $P_{SS}(x)$ exhibits a power-law
decay in $x$ only for values of $x$ that are much smaller than the sample
size $L$. Since the roughness of the steady-state interface of super-rough
models at length scales much smaller than $L$ is described by the local
exponent $\alpha_l$, we expect the steady-state spatial persistence
probability in such models to exhibit a power-law decay with exponent
$\theta_{SS} = 1 - \alpha_l$ for $x \ll L$. For example, the
one-dimensional Mullins-Herring model~\cite{mh} is super-rough with
$\alpha_g=3/2$ and $\alpha_l=1$. For this model, the above argument
suggests that the steady-state spatial persistence exponent $\theta_{SS}$
is equal to 0, which agrees with the exact result reported in
Ref.~\cite{maj1}.

An important feature of our investigation is the development of a 
scaling description of the effects of a finite system size and a 
finite sampling distance on the measured persistence probabilities. 
We have also shown that the dependence of the FIC persistence and 
survival probabilities on the reference level $H$ (in atomistic 
models) or the width $w$ of the band (in continuum models) used in 
the selection of the subset of sampling sites is described by a 
scaling form. These scaling descriptions would be useful in the
analysis of experimental and numerical data on fluctuations in 
spatially extended stochastic systems.

Some of the numerical results reported here (such as the behavior 
of the SS survival probability and the forms of the scaling 
functions that describe the dependence of the persistence 
probabilities on the parameters $L$, $\delta x$ and $H$ or $w$) 
should be amenable to analytic treatment, especially for the EW 
equation with white noise, whose spatial properties can be 
mapped~\cite{maj1} to the temporal properties of the well-known 
random walk problem. Further work along these lines would be very 
interesting. The spatial persistence and survival probabilities 
considered here should be measurable in imaging experiments on
step fluctuations~\cite{exp_dan,exp_dan2}. Such experimental 
investigations would be most welcome.

\vspace{0.5 cm}
{\bf ACKNOWLEDGMENTS}

This work is partially supported by the US-ONR, the LPS, and the 
NSF-DMR-MRSEC at the University of Maryland. The authors would like 
to thank Satya N. Majumdar for several useful discussions. 
M.C. acknowledges useful discussions with E.D. Williams and
D.B. Dougherty.



\end{document}